**TokenSE: a Mamba-based discrete token speech enhancement framework for cochlear implants**

Hsin-Tien Chiang,[1] and John H. L. Hansen[1,a]

[1] *Cochlear Implant Processing Laboratory–Center for Robust Speech Systems (CRSS-CILab), The University of Texas at Dallas, Richardson, Texas 75080, USA*

Speech enhancement (SE) is critical for improving speech intelligibility and quality in real-world environments, particularly for cochlear implant (CI) users who experience severe degradations in speech understanding under noisy and reverberant conditions. In this study, we propose TokenSE, a discrete token-based SE framework operating in the neural audio codec space, which predicts clean codec token indices from degraded speech using a Mamba-based model. Unlike the earlier Transformer architecture, whose self- attention mechanism has a computational complexity that grows quadratically with sequence length, the input-dependent selection mechanism of Mamba achieves linear complexity, making it a compelling alternative to Transformers, especially for CI and hearing-aid (HA) applications. Objective evaluations show that TokenSE consistently outperforms baseline methods on both in-domain and out-of-domain datasets. Moreover, subjective listening experiments with CI users indicate clear benefit in speech intelligibility under adverse noisy and reverberant environments.

---

[a] Email: john.hansen@utdallas.edu

## I. INTRODUCTION

In real-world environments, speech signals are often degraded by both background noise and room reverberation, introducing complex distortions that severely compromise speech quality and intelligibility. Such degradations not only negatively affect normal- hearing (NH) listeners, but their impact is even more severe for cochlear implant (CI) users, who have limited time-frequency (T-F) content decoding capabilities, leading to a significant reduction of sentence-level and word-level speech understanding (Kokkinakis *et al.*, 2011; Hazrati *et al.*, 2013a; Hazrati and Loizou, 2013; Ren *et al.*, 2018). This is due to the masking of spectro-temporal cues, blurring of formant transitions, and a reduction of perceived envelope structure (amplitude modulations), all of which collectively and individually impair CI speech understanding in reverberation and noise co-exist environments (Hazrati *et al.*, 2013b; Chiang and Hansen, 2025). As a result, developing effective speech enhancement (SE) methods which have the potential to transition to CI/HA platforms in the future is important.

The rapid advancement of deep learning (DL) has shown effectiveness in suppressing noise and reverberation, achieving superior results over traditional statistical based approaches (Loizou, 2007). Common approaches in SE is to use discriminative models that learn to directly map noisy speech to the corresponding clean speech target. These approaches are broadly categorized into both T-F domain and time-domain approaches. In T-F domain methods, speech signals are transformed into a T-F representation using short-time Fourier transform (STFT). Enhancement methods then target this domain include masking-based approaches, such as the ideal binary mask (IBM) (Wang, 2005), ideal ratio mask (IRM) (Narayanan and Wang, 2013), and complex ratio mask (CRM) (Williamson *et al.*, 2015), as well as mapping-based approaches that directly estimate the spectral representation of clean speech (Lu *et al.*, 2013; Xu *et al.*, 2013). Alternatively, time- domain methods estimate the clean speech signal directly from the raw waveform (Fu *et al.*, 2017; Pandey *et al.*, 2019; Défossez *et al.*, 2020). However, since many of these supervised methods are trained on a finite dataset, they cannot generalize well to unseen noise types and reverberation in real-world environments. Additionally, some discriminative approaches can introduce distortions that outweigh the benefits of noise reduction (Wang *et al.*, 2019). Therefore, the ability to balance noise suppression while maintaining natural speech production properties as well as factors that ensure auditory perception benefits has been an ongoing challenge (Hansen and Clements, 1991; Hansen and Nandkumar, 1995).

Alternatively, there are SE approaches that utilize generative models to learn a prior distribution over clean speech. These models are aimed at learning the inherent characteristics of clean speech, including spectral and temporal structure, which then serve as prior knowledge for inferring clean speech from noisy or reverberant inputs. Various deep generative approaches have been explored, including generative adversarial networks (GANs) (Pascual *et al.*, 2017; Baby and Verhulst, 2019), variational autoencoders (VAEs) (Bando *et al.*, 2018; Leglaive *et al.*, 2018), flow-based models (Nugraha *et al.*, 2020), and diffusion-based generative models (Lu *et al.*, 2021; Richter *et al.*, 2023). However, under adverse conditions, such generative models are potentially prone to confusing phoneme structure, leading to the generation of unintended vocalizing artifacts (Richter *et al.*, 2023; Lemercier *et al.*, 2023).

Recently, the emergence of discrete representations extracted from neural audio codecs (NAC) has motivated their application to SE. In such frameworks, DL models predict the discrete acoustic tokens of clean speech from degraded inputs and reconstruct the enhanced waveform through a neural codec decoder. SELM (Wang *et al.*, 2024) regenerates noisy semantic tokens into clean versions using a language model (LM), while (Wang *et al.*, 2023; Yang *et al.*, 2024) predict clean token sequences conditioned on noisy inputs and auxiliary features extracted from self-supervised speech models, which provide semantic representations to assist token prediction. GenSE (Yao *et al.*, 2025) employs a hierarchical two-stage framework. In the first stage, an LM predicts the semantic tokens extracted from degraded speech to obtain enhanced semantic representations. In the second stage, a second LM jointly processes the enhanced semantic, degraded semantic, and acoustic tokens to generate enhanced acoustic tokens, which are then decoded to reconstruct the enhanced speech.

However, virtually all of these aforementioned methods have been developed specifically for NH individuals, with limited adaptation to improving speech understanding in CI users. Earlier attempts to address noise and reverberation in CIs primarily relied on signal- processing-based approaches. For example, a combination of speech pause detection and nonlinear spectral subtraction was shown to improve speech recognition under both steady-state and babble noise conditions (Yang and Fu, 2005). A SE method based on subspace principles has been proposed, showing improvements in sentence recognition scores under stationary noise conditions (Loizou *et al.*, 2005). A CI channel-selection method that exploited the signal-to-reverberant ratio of individual frequency channels achieved more than a 60% improvement in intelligibility under highly reverberant conditions (Kokkinakis *et al.*, 2011). In a later study, that approach was refined to eliminate reliance on prior knowledge of the room impulse response or the anechoic signal, resulting in an average performance gain of 32.21 percentage points (Hazrati and Loizou, 2013). A blind single-channel ratio-masking strategy, designed to jointly suppress reverberation and noise, yielded an average improvement of approximately 5% in speech intelligibility scores for CI users (Hazrati *et al.*, 2013b). In addition, a method combining harmonic structure estimation with the traditional minimum mean square error (MMSE) SE approach showed improvement in intelligibility from 2% to 16% in 0 dB babble noise (Wang and Hansen, 2018).

More recently, several DL methods have also been explored for CI users. A deep denoising autoencoder integrated with a noise classifier maps noisy log-power spectral features to clean speech and has been shown to improve intelligibility for Mandarin CI recipients in non-stationary noise environments (Lai *et al.*, 2018). SE models based on dual-path recurrent neural networks have been investigated for CI users to suppress noise and reverberation in both single- and multi-microphone settings, with multi-microphone configurations showing greater intelligibility improvements (Gaultier and Goehring, 2024). A deep complex convolutional transformer network was proposed for complex spectral mapping, simultaneously enhancing both magnitude and phase, and achieving a 40% improvement in speech intelligibility under noisy conditions (Mamun and Hansen, 2024). A GAN-based architecture incorporating deformable convolutional networks has also been proposed for reverberation suppression, resulting in a 29% gain in speech intelligibility (Chiang and Hansen, 2025). Subjective evaluation results suggest the promising potential of these two DL-based approaches for restoring intelligibility in CI users under adverse noisy or reverberant conditions.

Inspired by the large intelligibility gains achieved using DL-based methods, this current study proposes TokenSE, a discrete token–based SE framework operating in the neural audio codec token space. TokenSE uses a Mamba-based model to predict clean codec tokens from degraded speech inputs for CI users in both noisy and reverberant environments. Mamba (Gu and Dao, 2024) is a state-space model with a selective mechanism that has received attention for sequence modeling. It offers linear complexity in token length while reaching performance that matches or even surpasses Transformer models. In speech research, Mamba has been applied for automatic speech recognition (Zhang *et al.*, 2025; Jiang *et al.*, 2025a), speech separation (Li *et al.*, 2024a; Jiang *et al.*, 2025a,b), and SE (Chao *et al.*, 2024; Zhang *et al.*, 2025). However, its application to discrete token-based SE framework has not yet been explored, and such frameworks have especially not been investigated for the unique hearing decoding needs of CI users.

Our main contributions in this study are five-fold. (1) To the best of our knowledge, this work proposes the first discrete token- based SE framework for CI users. (2) TokenSE jointly addresses both noise and reverberation, whereas most existing studies focus on either noise (Wang and Hansen, 2018; Lai *et al.*, 2018; Mamun and Hansen, 2024) or reverberation (Kokkinakis *et al.*, 2011;

Hazrati and Loizou, 2013; Chiang and Hansen, 2025) in isolation, which underestimates the challenges encountered in real-world listening environments. (3) Within the TokenSE framework, we introduce Mamba for modeling clean codec token prediction in SE and systematically compare its performance with Transformer-based counterparts, as well as representative discriminative and generative baseline models. (4) Unlike prior works (Wang *et al.*, 2023; Yang *et al.*, 2024) that keep the NAC encoder fixed and rely on auxiliary features, we fine-tune the NAC encoder jointly within the TokenSE framework to better adapt token representation to the specific degraded speech conditions. (5) Subjective evaluations using CI listeners highlight the potential benefit of the proposed framework to significantly enhance speech intelligibility under adverse listening conditions.

The remainder of this paper is organized as follows. Section II introduces the proposed discrete-token-based SE framework. Section III and IV describe the experimental setup and the objective and subjective evaluation results. In Section V, we discuss the findings of this study. Finally, Section VI concludes the paper.

## II. METHODOLOGY

### A. Mamba

#### 1. From state space model to Mamba

The core of Mamba is a linear selective state space model (SSM). SSM has gained attention as an alternative to recurrent neural networks and Transformers due to its efficiency in capturing long-range dependencies (Gu *et al.*, 2025). Derived from continuous systems, SSM maps the input $x(t) \in R$ to the output $y(t) \in R$ at time through a hidden state $h(t) \in R^M$. The mapping process can be represented as follows:

$$h'(t) = Ah(t) + Bx(t) \qquad (1)$$

$$y(t) = Ch(t) \tag{2}$$

where $h'(t)$ denotes the derivative of the current state $h(t)$, $A \in R^{M \times M}$ is the state transition matrix that indicates how states change over time, $B \in R^{M \times 1}$ is the input matrix that determines how inputs affect state updates, and $C \in R^{1 \times M}$ is the output matrix that indicates how the output is generated from the current states.

To obtain the discrete form, a scaling parameter $\Delta$ is used to transform continuous matrices $A$, $B$ to their discrete counterparts $\bar{A}, \bar{B}$. This transformation commonly utilizes the zero-order hold (ZOH), defined as:

$$\bar{A} = exp(\Delta A) \tag{3}$$

$$\bar{B} = (\Delta A)^{-1}(exp(\Delta A) - I)\Delta B \tag{4}$$

With the discrete process into $\bar{A}, \bar{B}$, Eq. (1) and (2) can then be rewritten as:

$$h_t = \bar{A}h_{t-1} + \bar{B}x_t \tag{5}$$

$$y_t = Ch_t \tag{6}$$

Next, the output $y$ is then computed through a global convolution:

$$\bar{K} = (C\bar{B}, C\bar{A}\bar{B}, ..., C\bar{A}^{L-1}\bar{B}) \tag{7}$$

$$y_t = Ch_t \tag{8}$$

where $L$ is the length of the input sequence.

However, SSMs are time-invariant, meaning that their parameters are defined as time-independent. To address this limitation, Mamba introduces a selective mechanism that parameterizes $B, C$, and $\Delta$ as functions of the input $x$. This allows Mamba to selectively focus or actually ignore specific parts of the sequence, thereby enhancing adaptability and efficiency. The discretization process after incorporating the selection mechanism is as follows:

$$\bar{B} = s_B(x) \tag{9}$$

$$\bar{C} = s_C(x) \tag{10}$$

$$\Delta = \tau_\Delta(Parameter + s_\Delta(x)) \tag{11}$$

where $\bar{B} \in R^{B \times L \times M}$, $\bar{C} \in R^{B \times L \times M}$, and $\Delta \in R^{B \times L \times D}$. $s_B(x)$ and $s_C(x)$ are linear functions that project the input $x$ into a dimension $M$, while $s_\Delta(x)$ projects the hidden state dimension $D$ linearly into the desired dimension. Through this design, the discrete SSM has therefore changed from a time-invariant to a time-varying model. In addition, Mamba incorporates hardware-aware optimizations including parallel scan and recomputation to improve computational efficiency and minimize memory overhead. This suggests benefits for system operation in the CI field.

*2. Mamba variants*

The Mamba models which are investigated in this study include three distinct variants (see Figure 1). The first is the unidirectional Mamba, which we term Mamba (Uni). This architecture performs causal computations that rely solely on past information due to the inherent nature of SSMs. To overcome this limitation and enable the model to utilize information from both past and future contexts, we also employ the bidirectional Mamba, termed Mamba (Bi) (e.g., the second option in Figure 1). This is achieved by running two SSMs with causal convolutions in parallel, with one processing the original sequence and the other processing the reversed sequence. The outputs from the two directions are then concatenated and passed through a transposed convolution layer. Lastly, we investigate a hybrid variant inspired by the work of (Zhang *et al.*, 2025), which we term Transformer−MHSA+Mamba (Bi) (e.g., the third option in Figure 1). This model replaces the standard multi-head self-attention (MHSA) module within a Transformer with our bidirectional Mamba component while retaining the feed-forward network and layer normalization. In this design, SSMs perform efficient linear computations, whereas the remaining Transformer layers introduce nonlinearity. An overview of the three Mamba variants, including Mamba (Uni), Mamba (Bi), and Transformer−MHSA+Mamba (Bi), is illustrated in Figure. 1. Next, the neural audio codec used here is discussed.

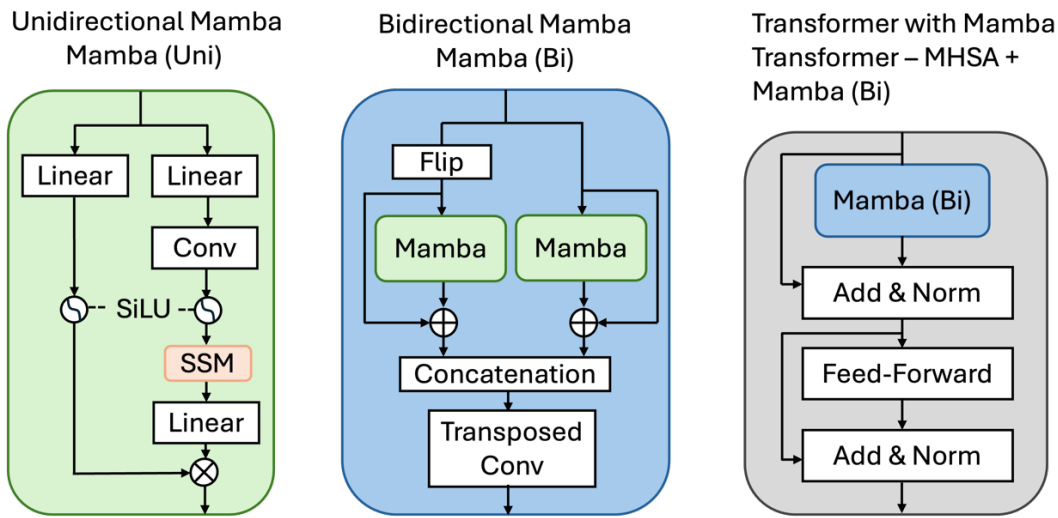

FIG. 1. (Color online). The illustrations of three Mamba variants: unidirectional Mamba (Mamba (Uni)), bidirectional Mamba (Mamba (Bi)), and a hybrid architecture replacing MHSA in Transformer layer with bidirectional Mamba (Transformer−MHSA+Mamba (Bi)).

**B. Neural audio codec (NAC)**

NAC (Zeghidour *et al.*, 2022; Défossez *et al.*, 2023; Yang *et al.*, 2023; Kumar *et al.*, 2024) performs encoding of audio signals into discrete codes, which are then used to reconstruct the original audio. The objective is to achieve efficient codec compression by using a minimum number of bits required for storage or transmission, while still maintaining high perceptual quality and avoiding any substantial degradation in the reconstructed audio process.

The general architecture of these end-to-end codec models comprises three core modules: an encoder, a Residual Vector Quantization (RVQ) module, and a decoder. These components are trained jointly, leveraging both reconstruction and adversarial objectives. The encoder compresses the input waveform to a much lower sampling rate, and these representations are then discretized using RVQ. In this process, the continuous bottleneck features are mapped into sequences of discrete code tokens through a cascade of VQs. The first quantizer maps the features to the nearest entries in its codebook, while each subsequent quantizer operates on the residual error left by the previous stage. By iteratively refining the quantization, RVQ achieves a more accurate discrete representation using multiple smaller codebooks. Finally, the decoder reconstructs the audio waveform by upsampling from these quantized features in the time domain.

Pioneering work in this area include such systems as SoundStream (Zeghidour *et al.*, 2021), which first introduced the RVQ module. Building upon Soundstream, Encodec (Défossez *et al.*, 2023) incorporated LSTM layers together with a small transformer-based language model over the quantized units to enhance sequential modeling. HiFi-Codec (Yang *et al.*, 2023) introduced group residual vector quantization, which reduces the number of codebooks required. Descript-audio-codec (DAC) (Kumar *et al.*, 2024) proposed several improvements over Encodec. First, it mitigates the codebook collapse problem by performing quantization in a very low-dimensional latent space. Second, it replaces the conventional ReLU activation with the Snake activation function, which has been shown to improve the reconstruction quality of periodic signals.

In this study, we adopt the Encodec as our NAC backbone. The encoder–decoder architecture is built on SEANet (Tagliasacchi *et al.*, 2020), where the $K$ RVQ layers are cascaded and each layer quantizes the latent features using a codebook of size 1024. We use the discrete tokens from the first four RVQ codebooks ($K = 4$) as the prediction targets.

**C. Proposed overall framework**

As shown in Figure 2, the proposed TokenSE framework employs Mamba modules to predict clean speech tokens from degraded inputs. Within TokenSE, the degraded waveform is first compressed into a latent embedding $Z_d$ using the Encodec encoder. Taking $Z_d$ as input, the Mamba modules are trained to output the predicted code tokens $\hat{c}$ that approximate the ground-truth clean tokens $c$. Each predicted token corresponds to a discrete index in the learned codebook, and mapping these indices to their entries produces the quantized representation $\hat{Z}_c$. The quantized features are then passed through the Encodec decoder to reconstruct the enhanced waveform. In this framework, the decoder and codebooks are kept fixed, while the encoder and Mamba modules are jointly optimized. This design choice is because the pretrained codec encoder is not explicitly optimized for SE tasks, which may result in a domain mismatch when processing degraded signals. By fine-tuning the encoder together with the Mamba modules, the framework can learn enhancement-specific acoustic representations directly from the degraded waveform.

TokenSE consists of 12 Mamba layers followed by a linear layer. Each Mamba layer has a model dimension of $d_{model} = 256$, an SSM state dimension of $d_{SSM} = 16$, a local convolution width of $d_{conv} = 4$, and an expansion factor of $E = 2$. The linear layer predicts the probability distribution over token indices. For token generation, tokens within each quantizer are predicted in parallel, while across quantizers the process follows the hierarchical design of RVQ, where each quantizer conditions on the outputs of all preceding quantizers.

To train TokenSE, we define the loss function $L$, which consists of two components: (1) code token prediction loss $L^{token}$, formulated as a cross-entropy loss to supervise the code token prediction, and (2) code entry reconstruction loss $L^{entry}$, formulated as an L2 loss that forces the retrieved entries $\hat{Z}_c$ obtained by the predicted tokens $\hat{c}$ to approach the codebook entries $Z_c$ of the clean speech. The total loss is defined as:

$$L = \lambda L^{token} + L^{entry} \tag{12}$$

$$L^{token} = -\frac{1}{N}\sum_{i=1}^{N} c_i \log(\hat{c}_i) \tag{13}$$

$$L^{entry} = \frac{1}{ND}\sum_{i=1}^{N}\sum_{j=1}^{D}(\hat{Z}_c - sg(Z_c))^2 \tag{14}$$

where $\lambda$ is set to 0.5, $N$ is the number of the code tokens, $D$ corresponds to the dimension of each codebook entry, and $sg(\cdot)$ denotes the stop-gradient operator.

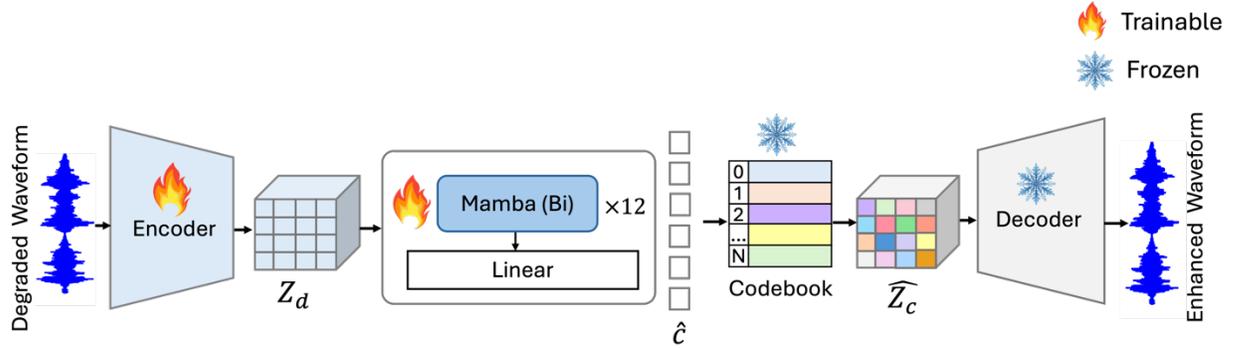

FIG. 2. (Color online). Overall framework of the proposed TokenSE method. TokenSE employs a discrete-token-based generative enhancement strategy, where the encoder extracts latent embeddings, Mamba modules predict clean code tokens, and the fixed codebooks and decoder reconstruct the enhanced waveform.

## III.    EXPERIMENTAL SETUP

### A.  Dataset

Two speech datasets are used in this study to evaluate both in-domain performance and out-of-domain (OOD) generalization. The in-domain dataset is derived from the DNS Interspeech 2020 Challenge (Reddy *et al.*, 2020) and is used for model training and benchmarking against existing approaches. The OOD dataset is constructed from the TIMIT corpus (Garofolo *et al.*, 1988) and is used to assess generalization under unseen acoustic conditions and to conduct subjective listening evaluations. While other corpus such as AzBio (Spahr *et al.*, 2012) and IEEE sentences (Rothauser, 1969) have often been used for CI testing, here we use DNS Challenge and TIMIT in order to compare with other SE studies. We rely on CI subject testing to demonstrate system performance for CI users.

#### *1.  In-domain dataset*

The clean speech set comes from the Librivox audiobook dataset, which contains over 500 hours of speech from 2,150 speakers. The noise dataset includes over 181 hours of clips from 150 classes which is sourced from Audioset (Gemmeke *et al.*, 2017) and Freesound. Room impulse responses (RIRs) are taken from the openSLR26 and openSLR28 datasets (Ko *et al.*, 2017), which include 3,076 real RIRs and approximately 115,000 synthetic RIRs. These RIRs are randomly selected to simulate reverberant conditions. Seventy-five percent of the clean utterances were convolved with randomly selected RIRs, and noisy speech was dynamically generated by mixing either clean or reverberant signals with noise at randomly chosen signal-to-noise ratios (SNRs) between −5 and 20 dB. We randomly select 90% of the utterances for training and 10% for validation.

For evaluation, the DNS Challenge provides a publicly available test dataset, including both synthetic and real recorded test sets. The synthetic set consists of two subsets: 150 clips without reverberation (Without Reverb) and 150 clips with reverberation (With Reverb), each spanning SNRs from 0 to 20 dB. The real recorded test set (Real Recordings) contains 300 audio clips.

*2. Out-of-domain (OOD) dataset*

The OOD dataset was primarily designed for controlled subjective evaluation and additionally used to assess model generalization under unseen acoustic conditions. To ensure a clear domain separation, the clean speech, noise, and RIR sources are all distinct from those used in the in-domain dataset. Eighty clean utterances were randomly selected from the TIMIT testing set to construct the dataset. Noise samples were taken from the NOISEX-92 database (Varga *et al.*, 1992), covering all fifteen noise types, and two RIRs were obtained from the REVERB Challenge corpus (Kinoshita *et al.*, 2013) with reverberation times ($T_{60}$) of 0.5 s and 0.7 s. Following a setup consistent with the in-domain evaluation, the OOD dataset includes both without-reverberation (noisy-only) and with-reverberation (reverberant + noisy) conditions. Specifically, the without-reverberation condition includes noisy-only speech at 0 and 5 dB SNR, while the with-reverberation condition comprises reverberant–noisy speech generated by first convolving clean utterances with RIRs of $T_{60} = 0.5$ s and 0.7 s, and then mixing them with noise at 5 dB SNR (Hazrati and Loizou, 2012).

**B. Evaluation metrics**

Objective evaluations are conducted on both the in-domain and OOD datasets to assess model performance and generalization capability. In line with previous studies (Wang *et al.*, 2024; Yang *et al.*, 2024; Yao *et al.*, 2025), we employ the DNSMOS P.835 model (Reddy *et al.*, 2022), as reference-based metrics such as PESQ are not reliable when temporal misalignment exists between the reference and enhanced speech. DNSMOS P.835 is a reference-free metric that estimates three perceptual dimensions: speech quality (SIG), background noise quality (BAK), and overall quality (OVR), where higher scores indicate better perceived quality.

For subjective evaluation, the mean opinion score (MOS) is used to assess perceived speech quality. Participants rate each stimulus on a five-point scale, where a score of 1 corresponds to poor quality and a score of 5 corresponds to excellent quality. Speech intelligibility is evaluated using the word recognition rate (WRR), which is computed based on the test samples.

**C. Subjective evaluation with CI listeners**

*1. Participant demographics*

Six adult CI users participated in this study, comprising three males and three females. Five participants were post-lingually deafened, while one was pre-lingually deafened. Participants' ages ranged from 20 to 73 years, with a mean age of 53.5 years. The duration of cochlear implant use varied from 1.25 to 18 years. All participants were native English speakers and were implanted with a variety of cochlear implant systems from different manufacturers. Table I provides the demographic information for all participants. Subjects were paid for their participation in this study.

TABLE I. Demographic information of CI recipients for subjective evaluation.

|  |  | S1 | S2 | S3 | S4 | S5 | S6 |
|---|---|---|---|---|---|---|---|
|  |  | Right/Left | Right/Left | Right/Left | Right/Left | Right/Left | Right/Left |
| Implant Specific | Implant type | CI622/HiRes 90K | CI624 | CI512/CI612 | CI522/CI622 | CI24RE/CI422 | Mi1200 |
|  | Uni/Bilateral | B | B | B | B | B | B |
|  | Speech Processor | CP1000/Naida CI M90 | CP1110 | CP1170/CP1110 | N/A | CP1150 | Me1510 |
|  | Device Exp. (years) | 3.5/18 | 1.25/2.5 | 8/1.5 | 7/4 | 14/10.5 | 12/9 |
| Processor Specific | Active electrodes | 22/16 | 22 | 22/21 | 22/N/A | 22 | 20 |
|  | Stim. Rate (Hz) | 900/N/A | 500 | 900/500 | N/A | 900 | 900 |
|  | 'n-maxima' | 8/N/A | 8 | 8/14 | N/A | 8 | 8 |
| Demographics | Age (years) | 20 | 73 | 72 | 68 | 58 | 30 |
|  | Linguistic Exp. | Pre-lingual | Post-lingual | Post-lingual | Post-lingual | Post-lingual | Post-lingual |

*2. Stimuli and procedure*

The stimuli used in the subjective evaluation were selected from the OOD dataset. The root-mean-square (RMS) level of all utterances was normalized to approximately 65 dB, and all audio files were sampled at 16 kHz. The listening tests were conducted using a graphical user interface (GUI) implemented in Python. CI participants performed the tests while fitted with their daily clinical processors.

For the speech intelligibility task, 80 stimuli were generated by crossing four acoustic scenarios with four processing methods. As described in Section III.A.2, the acoustic scenarios included two without-reverberation conditions (0 dB and 5 dB SNR) and two with-reverberation conditions ($T_{60} = 0.5$ s and 0.7 s, both at 5 dB SNR). The processing methods consisted of the unprocessed mixture, a signal-processing-based baseline, TokenSE, and the clean reference. Each scenario–method combination included five unique utterances, resulting in a total of 80 stimuli (4 scenarios × 4 methods × 5 utterances). Participants were asked to type the words they heard into a designated box within the Python GUI. The presentation order was fully randomized for each participant.

For the speech quality task, participants first listened to the clean speech as a reference, followed by three processed samples presented in random order for rating. The test sets were constructed using the same four acoustic scenarios as in the intelligibility task, with five utterances per scenario, resulting in a total of 20 test sets. As the clean reference was not rated, each participant evaluated 60 stimuli in total (4 scenarios × 5 test sets per scenario × 3 processing methods per set).

**D. Implementation details**

During training, we randomly cut a 4-second segment for each training utterance. TokenSE is trained using four 2080 GPUs with a batch size of 16. We use the Adam optimizer with an initial learning rate of $2 \times 10^{-4}$, and early stopping is applied to finish training if there is no improvement in validation set for 20 consecutive epochs.

**IV. RESULTS**

**A. In-domain evaluation**

We compared TokenSE with one signal-processing-based method and four DL-based approaches: two discriminative models and two generative models. The signal-processing baseline is based on the Logarithmic Minimum Mean Square Error (Log-MMSE) (Ephraim and Malah, 2003) estimator, which minimizes the mean-square error of the log-spectral amplitude. It is selected given that CI processing primarily rely on traditional signal-processing methods. The two discriminative models include DEMUCS (Défossez *et al.*, 2020), which employs a multi-layer convolutional encoder-decoder architecture with U-Net skip connections and a sequence modeling module, and FRCRN (Li *et al.*, 2024b), a convolutional recurrent encoder-decoder network designed to enhance feature representations through frequency recurrence. Generative models include SELM (Wang *et al.*, 2024) and MaskSR (Li *et al.*, 2024b), with small (MaskSR-S) and medium (MaskSR-M) variants. Both models leverage language models (LMs) to predict token sequences for re-synthesizing enhanced speech. Specifically, SELM transforms noisy speech into discrete tokens obtained from a pretrained SSL model through k- means clustering, and then employs LM to generate clean-speech tokens, which are decoded back into waveform using a neural vocoder. MaskSR operates in the NAC token space based on DAC (Kumar *et al.*, 2024), where a masked LM is trained to predict the masked acoustic tokens of clean speech conditioned on the encoded corrupted input. The predicted tokens are subsequently passed through the decoder to reconstruct the enhanced waveform. We also compare TokenSE by replacing its Mamba-based backbone with a Transformer-based counterpart, where the Transformer uses a model dimension of 256 and eight attention heads.

Table II presents the performance results on the DNS Challenge test sets. First, we compare TokenSE with different backbone architectures, including the non-causal Transformer and two Mamba-based variants: Mamba (Bi) and Transformer-MHSA + Mamba (Bi). The results show that both Mamba-based variants consistently outperform the Transformer counterpart. While both achieve comparable results, replacing MHSA with Mamba performs best on the reverberant dataset, whereas using Mamba alone yields higher scores on the non-reverberant and real recordings. Given that replacing MHSA with Mamba introduces more parameters than directly using Mamba, we select Mamba (Bi) as the backbone for TokenSE due to its favorable balance between performance and parameter efficiency. This configuration is used for all subsequent comparisons and further experiments.

For the causal version of TokenSE, we compare a causal Transformer and Mamba (Uni) as backbones. Mamba (Uni) consistently outperforms the causal Transformer on both simulated and real datasets, even without access to future information. Although Mamba (Uni) shows slightly degraded performance compared to its bidirectional counterpart, Mamba (Bi), it demonstrates strong potential for real-time implementation, which is particularly beneficial for CI applications.

Finally, we compare TokenSE (with the Mamba (Bi) backbone) against baseline methods. The Log-MMSE approach provides limited improvement over the unprocessed speech but remains inferior to DL-based methods. In contrast, TokenSE consistently outperforms all discriminative and generative baselines across both simulated and real recordings, confirming its effectiveness.

TABLE II. Objective in-domain evaluation results on the DNS Challenge test set.

| System | With Reverb | | | Without Reverb | | | Real Recordings | | |
|---|---|---|---|---|---|---|---|---|---|
| | SIG | BAK | OVL | SIG | BAK | OVL | SIG | BAK | OVL |
| Unprocessed | 1.760 | 1.497 | 1.392 | 3.392 | 2.618 | 2.483 | 3.053 | 2.509 | 2.255 |
| Log-MMSE | 2.270 | 2.259 | 1.788 | 3.261 | 3.028 | 2.548 | 3.153 | 2.869 | 2.424 |
| DEMUCS [15] | 2.876 | 3.789 | 2.549 | 3.534 | 4.146 | 3.307 | 3.227 | 3.986 | 2.946 |
| FRCRN [56] | 2.934 | 2.924 | 2.279 | 3.574 | 4.154 | 3.331 | 3.371 | 3.978 | 3.037 |
| SELM [25] | 3.160 | 3.577 | 2.695 | 3.508 | 4.096 | 3.258 | 3.591 | 3.435 | 3.124 |
| MaskSR-S [57] | 3.524 | 4.016 | 3.223 | 3.575 | 4.082 | 3.307 | 3.398 | 4.011 | 3.103 |
| MaskSR-M [57] | 3.531 | 4.065 | 3.253 | 3.586 | 4.116 | 3.339 | 3.430 | 4.025 | 3.136 |
| **TokenSE (Proposed)** | | | | | | | | | |
| Transformer (causal) | 3.431 | 3.991 | 3.118 | 3.554 | 4.070 | 3.281 | 3.137 | 3.740 | 2.743 |
| Mamba (Uni) | 3.449 | 4.057 | 3.166 | 3.560 | 4.138 | 3.319 | 3.329 | 4.018 | 3.049 |
| Transformer | 3.585 | 4.076 | 3.305 | 3.601 | 4.150 | 3.363 | 3.358 | 4.026 | 3.072 |
| Transformer–MHSA + Mamba (Bi) | **3.643** | **4.173** | **3.403** | 3.629 | 4.180 | 3.407 | 3.483 | 4.110 | 3.219 |
| Mamba (Bi) | 3.630 | 4.164 | 3.394 | **3.650** | **4.187** | **3.431** | **3.490** | **4.120** | **3.233** |

## B. OOD evaluation

Following the subjective evaluation described in Section III.C.2, the OOD comparison focuses on the selected TokenSE configuration (with the Mamba (Bi) backbone) and Log-MMSE baseline. Both methods are directly evaluated on the OOD dataset to assess their generalization under unseen acoustic conditions.

### 1. Objective evaluation

Tables III and IV present the objective OOD results under the without-reverberation (noisy-only) and with-reverberation (reverberant + noisy) conditions, respectively. Under the without-reverberation condition, both enhancement methods improve speech quality compared to the unprocessed input, with TokenSE consistently outperforming the Log-MMSE baseline at both 0 dB and 5 dB SNR. The improvement is more pronounced at 0 dB SNR, where the input is more severely degraded, while the smaller gain at 5 dB SNR reflects the reduced enhancement margin for relatively cleaner speech. Under the with-reverberation condition, TokenSE again achieves superior performance over both the unprocessed signal and the Log-MMSE baseline. A slight performance degradation is observed as the reverberation time increases from $T_{60} = 0.5$ s to 0.7 s, indicating the increased difficulty posed by stronger reverberation. Overall, these results demonstrate that TokenSE generalizes well to unseen acoustic conditions involving both additive noise and reverberation.

TABLE III. Objective OOD results under the without-reverberation (noisy-only) condition at 0 dB and 5 dB SNR.

| System | 0 dB SNR | | | 5 dB SNR | | |
| --- | --- | --- | --- | --- | --- | --- |
| | SIG | BAK | OVL | SIG | BAK | OVL |
| Unprocessed | 2.756 | 2.301 | 2.058 | 3.106 | 2.451 | 2.224 |
| Log-MMSE | 3.026 | 3.13 | 2.403 | 3.128 | 3.334 | 2.506 |
| TokenSE | **3.514** | **4.136** | **3.24** | **3.486** | **4.119** | **3.198** |

TABLE IV. Objective OOD results under the with-reverberation (reverberation + noisy) condition with $T_{60} = 0.5$ s and 0.7 s at 5 dB SNR.

| System | T60 = 0.5 s | | | T60 = 0.7 s | | |
| --- | --- | --- | --- | --- | --- | --- |
| | SIG | BAK | OVL | SIG | BAK | OVL |
| Unprocessed | 2.081 | 1.716 | 1.609 | 2.328 | 1.853 | 1.722 |
| Log-MMSE | 2.618 | 2.644 | 2.078 | 2.792 | 2.818 | 2.214 |
| TokenSE | **3.505** | **4.103** | **3.218** | **3.454** | **4.082** | **3.161** |

*2. Subjective evaluation*

a. *Without reverberation.* Figure 3 presents the mean WRR for CI recipients under the without-reverberation (noisy-only) condition at 0 dB and 5 dB SNR. At 0 dB SNR, the unprocessed mixture results in a very low WRR, indicating severe intelligibility degradation under highly adverse noise conditions. While Log-MMSE provides a moderate improvement, TokenSE yields a substantial intelligibility gain, improving WRR by 47.19 percentage points relative to the unprocessed signal. At 5 dB SNR, Log-MMSE slightly degrades intelligibility relative to the unprocessed input, whereas TokenSE consistently improves WRR by 30.21 percentage points and achieves performance close to the clean condition. A repeated-measures analysis of variance (ANOVA) was conducted on listener performance with a significance level of 0.05. The ANOVA results indicate a statistically significant difference across processing conditions at both 0 dB and 5 dB SNR, with $F(3, 15) = 46.31, p < 1 \times 10^{-7}$ and $F(3, 15) = 12.32, p < 2.5 \times 10^{-4}$, respectively. Post-hoc pairwise comparisons further reveal that, at 0 dB SNR, TokenSE shows a statistically significant improvement over both the unprocessed mixture ($p = 0.006$) and the Log-MMSE baseline ($p = 0.026$), while no statistically significant difference is observed relative to the clean reference condition ($p = 0.066$). At 5 dB SNR, TokenSE achieves a statistically significant improvement over the Log-MMSE baseline ($p = 0.015$), with no statistically significant differences observed relative to the unprocessed mixture ($p = 0.098$) or the clean condition ($p = 1.000$).

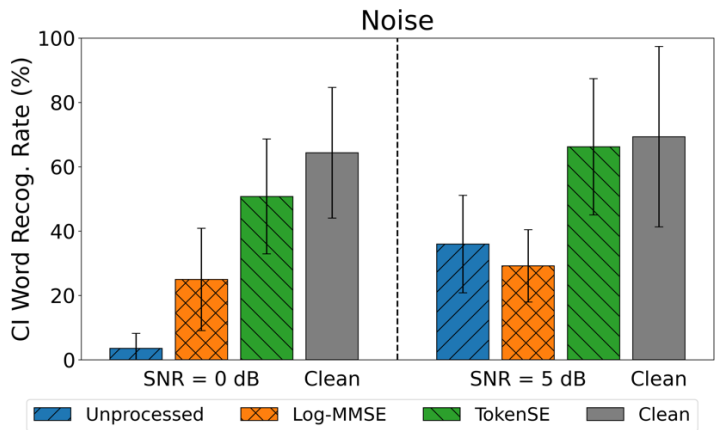

FIG. 3. (Color online). Mean word recognition rate under the without-reverberation (noisy-only) condition for CI recipients.

Figure 4 presents the MOS results. Across both SNR levels, consistent trends are observed, where the Log-MMSE baseline improves perceived speech quality relative to the unprocessed mixture, while TokenSE achieves the highest MOS among all processing conditions. The ANOVA results indicate a statistically significant difference across processing conditions at both 0 dB and 5 dB SNR, with $F(2,10) = 21.56, p < 2.36 \times 10^{-4}$ and $F(2,10) = 10.85, p < 3.126 \times 10^{-3}$, respectively. Post-hoc pairwise comparisons indicate that, at both SNR levels, TokenSE achieves a statistically significant improvement over the unprocessed mixture ($p = 0.0045$ at 0 dB and $p = 0.034$ at 5 dB), while exhibiting no statistically significant difference over the Log-MMSE baseline ($p = 0.086$ at 0 dB and $p = 0.234$ at 5 dB).

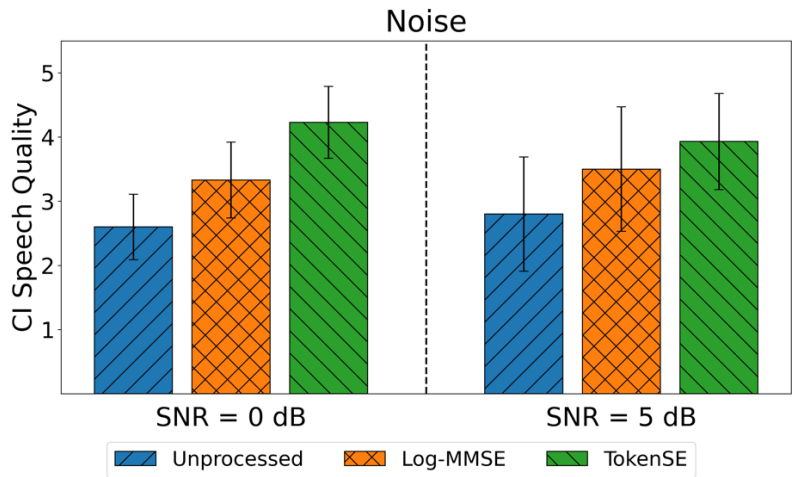

FIG. 4. (Color online). Mean speech quality scores under the without-reverberation (noisy-only) condition for CI recipients.

b. *With reverberation.* Figure 5 presents the mean WRR for CI recipients under the with-reverberation (reverberant + noisy) condition with $T_{60} = 0.5$ s and 0.7 s. The Log-MMSE baseline yields a slight intelligibility improvement at $T_{60} = 0.5$ s, but leads to a performance degradation when the reverberation time increases to $T_{60} = 0.7$ s. In contrast, TokenSE consistently improves intelligibility across both reverberation conditions, achieving WRR gains of 38.40 and 38.41 percentage points relative to the unprocessed mixture at $T_{60} = 0.5$ s and 0.7 s, respectively. The ANOVA results indicate a statistically significant difference across processing conditions at both $T_{60} = 0.5$ s and 0.7 s, with $F(3,15) = 40.24, p <$ $2.1 \times 10^{-7}$ and $F(3,15) = 29.51, p < 2 \times 10^{-6}$, respectively. Post-hoc pairwise comparisons further reveal that, at $T_{60} = 0.5$ s, TokenSE achieves statistically significant improvements over both the unprocessed mixture ($p = 0.013$) and the Log-MMSE baseline ($p = 0.00045$), while no statistically significant difference is observed relative to the clean reference condition ($p = 1.000$). A similar trend is observed at $T_{60} = 0.7$ s, where TokenSE yields statistically significant improvements over the unprocessed mixture ($p = 0.035$) and the Log-MMSE baseline ($p = 0.014$), again showing no statistically significant difference compared to the clean condition ($p = 1.000$).

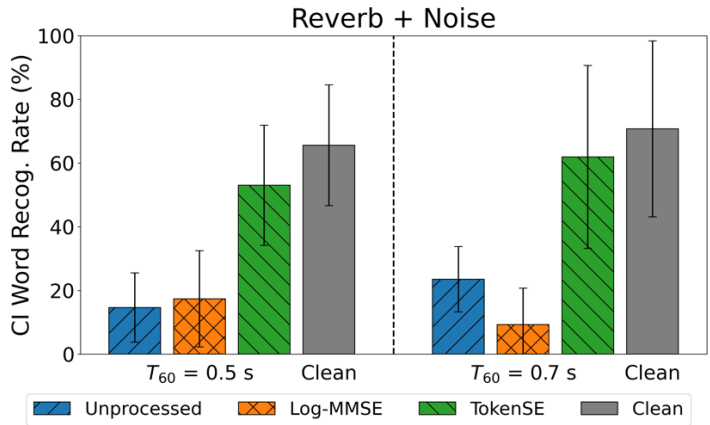

FIG. 5. (Color online). Mean word recognition rate under the with-reverberation (reverberant + noisy) condition for CI recipients.

Figure 6 presents the MOS results. Consistent trends are observed, where the Log-MMSE baseline improves perceived speech quality relative to the unprocessed mixture, while TokenSE achieves the highest MOS. The ANOVA results indicate a statistically significant difference across processing conditions at both $T_{60} = 0.5$ s and $0.7$ s, with $F(2, 10) = 10.52, p < 3.47 \times 10^{-3}$ and $F(2, 10) = 26.32, p < 1.04 \times 10^{-4}$, respectively. Post-hoc pairwise comparisons indicate that, at $T_{60} = 0.5$ s, no statistically significant differences are found between TokenSE and either the unprocessed mixture ($p = 0.0672$) or the Log-MMSE baseline ($p = 0.0539$). At $T_{60} = 0.7$ s, however, TokenSE exhibits statistically significant improvements compared with both the unprocessed mixture ($p = 0.00083$) and the Log-MMSE baseline ($p = 0.0098$).

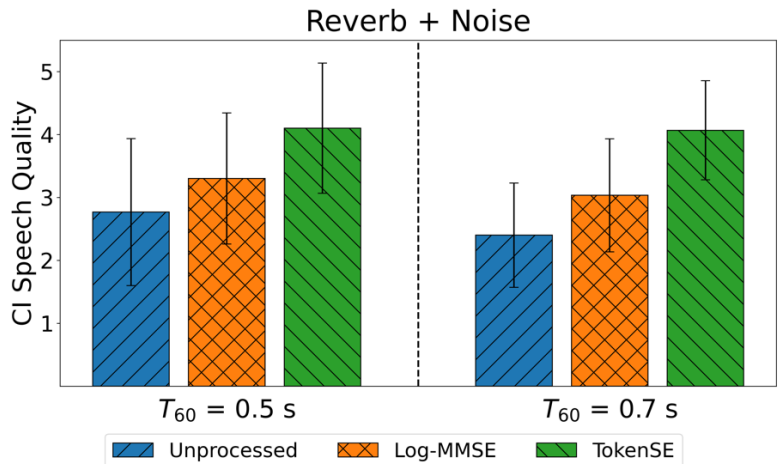

FIG. 6. (Color online). Mean speech quality scores under the with-reverberation (reverberant + noisy) condition for CI recipients.

### C. Ablation study

#### 1. Effect of encoder freezing and using auxiliary inputs in TokenSE

Previous studies (Wang *et al.*, 2023; Yang *et al.*, 2024) kept the encoder fixed while adding auxiliary input features together with encoder representations to predict clean token sequences. Following a similar setup, we also freeze the encoder and include auxiliary features in TokenSE framework. We explore two types of auxiliary features: Mel-spectrograms and the weighted sum of representations from the WavLM (Chen *et al.*, 2022) large model (WavLM-WS). The auxiliary input is first passed through a series of convolutional layers, each followed by ReLU activation and layer normalization. The resulting embedding is then linearly interpolated to match the token sequence length and passed through a linear layer before being added to the encoder representation.

As shown in Table V, freezing the pretrained encoder without auxiliary input results in inferior performance. Adding auxiliary inputs improves results, with WavLM-WS outperforming Mel-spectrograms on both simulated and real data. However, this improvement comes at the cost of increased computational complexity for the WavLM model compared to Mel-spectrograms processing. Overall, our results demonstrate that jointly fine-tuning the encoder within the TokenSE framework, as reported in Table II, consistently achieves better performance than freezing the encoder while adding auxiliary inputs, except for the SIG score, which remains comparable when using WavLM-WS on real recordings. Moreover, fine-tuning encoder eliminates the need for auxiliary feature processing, thereby reducing both the number of parameters and the overall computational complexity.

TABLE V. Performance comparison on the in-domain dataset under encoder-freezing conditions with different auxiliary inputs within the TokenSE framework.

| Fix Encoder | Auxiliary Input | With Reverb | | | Without Reverb | | | Real Recordings | | |
|---|---|---|---|---|---|---|---|---|---|---|
| | | SIG | BAK | OVL | SIG | BAK | OVL | SIG | BAK | OVL |
| ✓ | - | 3.529 | 4.134 | 3.281 | 3.576 | 4.159 | 3.347 | 3.41 | 4.014 | 3.157 |
| ✓ | Mel-spectrograms | 3.618 | 4.136 | 3.358 | 3.63 | 4.176 | 3.399 | 3.469 | 4.1 | 3.199 |
| ✓ | WavLM-WS | 3.627 | 4.416 | 3.373 | 3.641 | 4.161 | 3.405 | 3.49 | 4.095 | 3.229 |

*2. Computational complexity*

We analyze the computational complexity of backbone architectures used in TokenSE, including Transformer and Mamba (Bi), by estimating the number of floating-point operations (FLOPs). Figure 7 illustrates the GFLOPs of both models at different sequence lengths. As shown in the figure, Mamba (Bi) consistently performs less GFLOPs compared to Transformer. As the sequence length increases, the computational gap between Mamba (Bi) and Transformer becomes wider. This further highlights the computational efficiency of Mamba (Bi) compared to Transformer, making it a suitable backbone for the TokenSE framework.

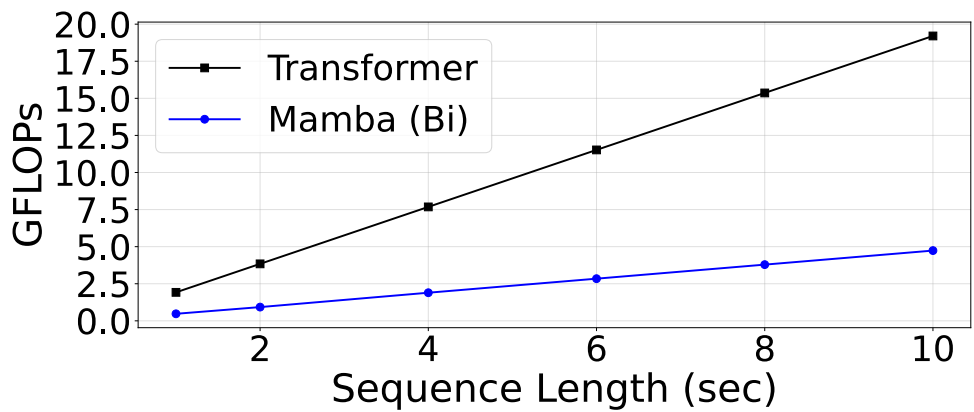

FIG. 7. (Color online). The number of FLOPs against the length of sequence between Transformer and Mamba (Bi).

## V. DISCUSSION

Figure 8 compares spectrograms of anechoic clean speech, unprocessed noisy-reverberant speech, speech enhanced by the Log-MMSE baseline, and speech processed by TokenSE. The comparison highlights the effectiveness of TokenSE in suppressing both reverberation and noise while preserving speech structure. In the clean spectrogram, clear phoneme boundaries and temporal gaps are preserved, whereas the unprocessed spectrogram exhibits pronounced spectral smearing caused by reverberation, along with widespread additive noise. As a result, temporal gaps are filled with residual noise and reverberation energy, obscuring phoneme onsets and offsets and masking critical word recognition cues that are essential for CI users (Hazrati and Loizou, 2022). The Log-MMSE baseline exhibits over-suppression, where substantial speech energy, particularly in the mid-to-high frequency bands, is excessively attenuated together with the background noise. Residual spectral smearing also remains evident, leading to blurred temporal structures and poorly defined word boundaries. This combination of over-suppression and residual smearing is particularly detrimental for CI users with limited spectro-temporal resolution, as it degrades temporal envelope cues essential for word boundary identification. In contrast, TokenSE effectively suppresses noise and reverberation while better preserving speech structure, restoring the envelope structure and preserving temporal transition boundaries, which increases the potential for CI users to perceive important transient segments of speech necessary to achieve high performance for sentence intelligibility.

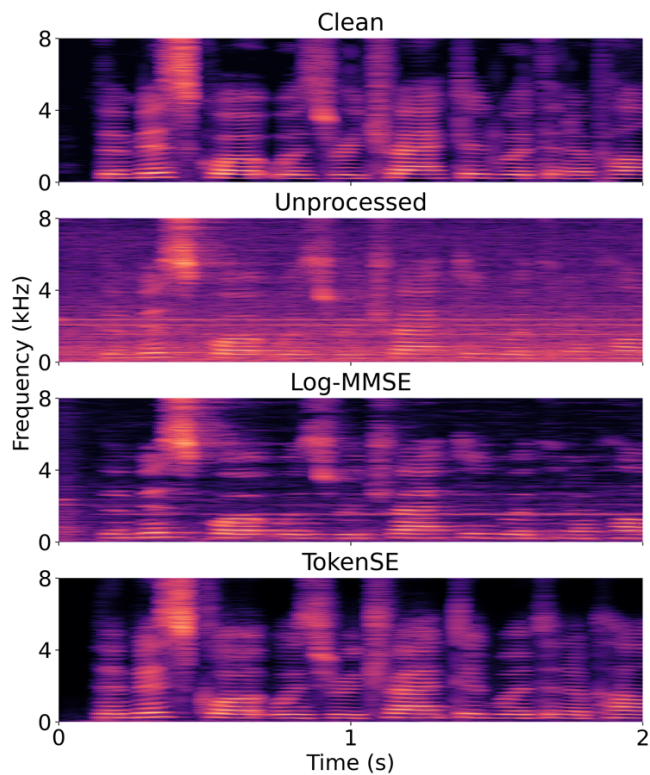

FIG. 8. (Color online). Spectrogram comparison of clean speech, unprocessed speech, and speech processed by Log-MMSE, and TokenSE under the with-reverberation (reverberant + noisy) condition ($T_{60}$ = 0.7 s, SNR = 5 dB). The spectrogram shows a segment of the utterance "The misquote was retracted with an apology." TokenSE achieves higher perceptual quality (OVRL = 2.71) compared to the unprocessed speech (OVRL = 1.18) and the Log-MMSE baseline (OVRL = 1.83).

## VI. CONCLUSION

In this study, we proposed TokenSE, a discrete-token-based SE framework to jointly address additive noise and reverberation. TokenSE operates in the NAC token space and employs a Mamba-based architecture to predict clean codec tokens from corrupted speech inputs. The use of a Mamba-based architecture enables efficient modeling of long-range temporal dependencies in codec token sequences while offering favorable computational efficiency. To the best of our knowledge, this is the first work to investigate discrete token-based SE specifically for CI users. Objective evaluation results on both in-domain and OOD datasets validate the effectiveness and generalization capability of TokenSE. Formal subjective listening test results with CI listeners further confirmed the effectiveness of TokenSE in recovering clear speech envelopes, thereby enhancing the perception of transient segments critical for speech quality and intelligibility for CI users.

## ACKNOWLEDGMENTS

This work was supported by Grant No. R01 DC016839-02 (PI, Hansen) from the National Institute on Deafness and Other Communication Disorders, National Institutes of Health, and partially by the University of Texas at Dallas (UTDallas) from the Distinguished University Chair in Telecommunications Engineering held by J.H.L. Hansen. We also acknowledge and thank the Cochlear Implant subjects for their time and commitment to supporting the CI research community in participating in the listener evaluation for this study.

## AUTHOR DECLARATIONS

**Conflict of Interest**

The authors have no conflicts to disclose.

**Ethics Approval**

Data collection for this study was approved by the University of Texas at Dallas IRB (IRB 24-247/ IRB 24-248). All participants provided informed consent before beginning the experiment.

## DATA AVAILABILITY

The data that support the findings of this study are available from the corresponding author upon reasonable request.

## REFERENCES (BIBLIOGRAPHIC STYLE)

Richter, J., Welker, S., Lemercier, J.-M., Lay, B., and Gerkmann, T. (**2023**). "Speech enhancement and dereverberation with diffusion-based generative models," IEEE/ACM Trans. Audio Speech Lang. Process., **31**, 2351–2364.

Rothauser, E. H. (**1969**). "IEEE recommended practice for speech quality measurements," IEEE Trans. Audio Electroacoust., 17(3), 225-246.

Spahr, A. J., Dorman, M. F., Litvak, L. M., Van Wie, S., Gifford, R. H., Loizou, P. C., Loiselle, L.M., Oakes, T. and Cook, S. (**2012**). "Development and validation of the AzBio sentence lists," Ear and hearing, **33**(1), 112–117.

Tagliasacchi, M., Li, Y., Misiunas, K., and Roblek, D. (**2020**). "SEANet: A multi-modal speech enhancement network," in Proc. Interspeech, pp. 1126–1130.

Varga, A., Steeneken, H. J. M., Tomlinson, M., and Jones, D. (**1992**). "The NOISEX-92 study on the effect of additive noise on automatic speech recognition," DRA Speech Res. Unit, Worcestershire, U.K.

Wang, D. (**2005**). "On ideal binary mask as the computational goal of auditory scene analysis," in Speech separation by humans and machines, Boston, MA: Springer US, pp. 181–197.

Wang, D., and Hansen, J.H. (**2018**). "Speech enhancement for cochlear implant recipients," J. Acoust. Soc. Am. **143**(4), 2244–2254.

Wang, H., Yu, M., Zhang, H., Zhang, C., Xu, Z., Yang, M., Zhang, Y., and Yu, D. (**2023**). "Unifying robustness and fidelity: A comprehensive study of pretrained generative methods for speech enhancement in adverse conditions," arXiv, September 16, 2023, preprint, v1.

Wang, P., Tan, K., and Wang, D. (**2019**). "Bridging the gap between monaural speech enhancement and recognition with distortion-independent acoustic modeling," IEEE/ACM Trans. Audio Speech Lang. Process. **28**, 39–48.
33